# Tuneable Radiation Field Aided Quantum Spin Hall Phase in Bi$_2$Se$_3$ Thin Film

Partha Goswami [1, a] and Udai Prakash Tyagi [2,b]

[1,2] D.B.College, University of Delhi, Kalkaji, New Delhi-110019, India

[a]physicsgoswami@gmail.com, [b]uptyagi@yahoo.co.in

**Abstract.** We show fledgling quantum spin Hall phase by the normal incidence of near-infrared circularly polarized radiation field on Bismuth Selenide doped with magnetic impurities. For this purpose, we start with a low-energy two-dimensional, time-dependent Hamiltonian. The time dependence in the Hamiltonian arises due to the optical field describable by the associated gauge field. We make use of the Floquet theory in the high-frequency limit to investigate the system. The optical field tuneability leads to the emergence of the spin Hall phase, when intensity of the incident radiation is high, from the quantum anomalous Hall phase. Interestingly, the former phase is achievable here even in the presence of the magnetic impurities.

**Main Text**

We consider a thin film of Bi$_2$Se$_3$ together with magnetic impurities with normal parallel to the z crystal growth direction. We denote the thickness of the thin film (along z direction) as '$d$'. Accordingly, the corresponding Hamiltonian $H(\mathbf{k})$ given below contains constant terms and the z derivatives. In the basis $(\mid p1_{z,\uparrow}^{even}\rangle \mid p2_{z,\uparrow}^{odd}\rangle \mid p1_{z,\downarrow}^{even}\rangle \mid p2_{z,\downarrow}^{odd}\rangle)$ of the hybridized states of $p_z$ Se orbital (of even parity) and $p_z$ Bi orbital (of odd parity), the momentum space dimensionless model Hamiltonian[1-11] of the system could be written as $H(\mathbf{k}) = (\epsilon(k)\sigma_0 + M\sigma_z)\otimes\tau_0 + \vartheta(k)\sigma_0\otimes\tau_z + A_1\{(a_x k_x \sigma_x + a_y k_y \sigma_y) + \eta a_z k_z \sigma_z\}\otimes\tau_x$, where in the ket $\mid p_j{}^{even}_{z,\sigma}{}^{odd}\rangle$ the symbol $\sigma = \uparrow\downarrow$ stands for the real spin, $\mathbf{k} = (k_x, k_y)$, $\eta < 1$, $M$ is the exchange field from the magnetic dopants, $\sigma_{x,y,z}$, and $\tau_{x,y,z}$, respectively, are the Pauli matrices for the spin and the orbital degrees of freedom. If one wishes to work with a lattice model, the following replacements are necessary: $a_j k_j \to \sin(a_j k_j)$ and $(a_j^2 k_j^2) \to 2(1 - \cos(a_j k_j))$ where $j = (x, y, z)$, and $a_j$ is the lattice constant along $j$ direction. It may be mentioned that the lattice constants of bulk Bi$_2$Se$_3$ are in the basal plane $a$ = 4.14 Å and along the $c$-axis $c$ = 28.64 Å. Here $a_x = a_y = a$ and $a_z = c$. The energies $\epsilon(k) = \epsilon_0 - D_1 c^2 \partial_z^2 + D_2 a^2 k^2$, and $\vartheta(k) = \vartheta_0 + B_1 c^2 \partial_z^2 - B_2 a^2 k^2$, and $k^2 = (k_x^2 + k_y^2)$. Thus, it is easy to see that the coefficients $B_2$ and $D_2$ serve as the first neighbor hopping in a lattice model. We have made the Hamiltonian dimensionless by dividing every term in $H(\mathbf{k})$ by the first neighbor hopping $B_2 = 3.31$ eV[**1, 2**] which is the highest energy value. In this scheme we have $\epsilon_0 = -0.003, A_1 = 0.31, \eta = 0.16, B_1 = 0.18, B_2 = 1, M = 0.08, D_1 = 0.024$, and $D_2 = 0.34$ following the values of these quantities given in ref. [**1,2**].

In order to obtain surface state Hamiltonian ($H^{surface}(\mathbf{k})$) we make the replacement $c k_z \to -ic\,\partial_z$ and look for states localized within the surface of the form $exp(-i\kappa z)$. Under the open boundary condition (OBC), we seek such a value of $\kappa$ for which this exponential will be vanishingly small for z = $\pm \frac{W}{2}$ where W is the thickness of the film. From above we find that $H^{surface}(k,\kappa) = (\epsilon_1(k,\kappa) \sigma_0 + M\sigma_z)\otimes\tau_0 + \vartheta(k,\kappa) \sigma_0\otimes\tau_z + A_1\{(ak_x\sigma_x + ak_y\sigma_y) - \eta c\kappa\sigma_z\}\otimes\tau_x$, where $\epsilon(k,\kappa) = \epsilon_0 + D_1 c^2\kappa^2 + D_2 a^2 k^2$, and $\vartheta(k,\kappa) = \vartheta_0 - B_1 c^2\kappa^2 - B_2 a^2 k^2$. The eigenvalues ($\in_j$) of this matrix is given by the quartic $\in_j^4 + \gamma_3(k,\kappa) \in_j^3 + \gamma_2(k,\kappa) \in_j^2 + \gamma_1(k,\kappa) \in_j + \gamma_0(k,\kappa) = 0$ where

$$\gamma_0(k,\kappa) = (\eta A_1 c\kappa)^4 + 2(\eta A_1 c\kappa)^2((A_1 ak)^2 + \vartheta^2(k,\kappa) - \epsilon^2(k,\kappa) - M^2)$$
$$+ ((A_1 ak)^2 + \vartheta^2(k,\kappa))^2 - 2(A_1 ak)^2(\epsilon^2(k,\kappa) - M^2),$$

$$\gamma_1(k,\kappa) = 4((\eta A_1 c)^2 \kappa^2 + (A_1 ak)^2)\epsilon(k,\kappa) + 4(\vartheta^2(k,\kappa)\epsilon(k,\kappa) - \epsilon^3(k) + \epsilon(k,\kappa) M^2),$$

$$\gamma_2(k,\kappa) = -2((\eta A_1 c)^2 \kappa^2 + (A_1 ak)^2) - 2(\vartheta^2(k,\kappa) - 3\epsilon^2(k,\kappa) + M^2), \gamma_3(k,\kappa) = -4\epsilon(k,\kappa). \tag{1}$$

In view of the Ferrari's solution of a quartic equation, we find the roots as

$$\in_j (s,\sigma,k,\kappa) = \sigma\sqrt{\frac{\eta_0(k,\kappa)}{2} - \frac{\gamma_3(k,\kappa)}{4}} + s\left(b_0(k,\kappa) - \left(\frac{\eta_0(k,\kappa)}{2}\right) + \sigma c_0(k,\kappa)\sqrt{\frac{2}{\eta_0(k,\kappa)}}\right)^{\frac{1}{2}}, \tag{2}$$

where $j = 1,2,3,4$, $\sigma = \pm 1$ is the spin index and $s = \pm 1$ is the band-index. Since, the spin index $\sigma$ occurs twice in Eq. (2), the term $\sqrt{\frac{\eta_0(k,\kappa)}{2}}$ does not act like magnetic energy. The functions appring in (2) are given by

$$\eta_0(k,\kappa) = \frac{2b_0(k,\kappa)}{3} + (\Delta(k,\kappa) - \Delta_0(k,\kappa))^{\frac{1}{3}} - (\Delta(k,\kappa) + \Delta_0(k,\kappa))^{\frac{1}{3}},$$

$$\Delta_0(k,\kappa) = \left(\frac{b_0^3(k,\kappa)}{27} - \frac{b_0(k,\kappa)d_0(k,\kappa)}{3} - c_0^2(k,\kappa)\right),$$

$$\Delta(k,\kappa) = \left(\frac{2}{729}b_0^6 + \frac{4d_0^2 b_0^2}{27} + c_0^4 - \frac{d_0 b_0^4}{81} - \frac{2b_0^3}{27} + \frac{2c_0^2 b_0 d_0}{3} + \frac{d_0^3}{27}\right)^{1/2},$$

$$b_0(k,\kappa) = \left\{\frac{3\gamma_3^2(k,\kappa) - 8\gamma_2(k,\kappa)}{16}\right\}, c_0(k,\kappa) = \left\{\frac{-\gamma_3^3(k,\kappa) + 4\gamma_3(k,\kappa)\gamma_2(k,\kappa) - 8\gamma_1(k,\kappa)}{32}\right\}$$

$$d_0(k,\kappa) = \frac{-3\gamma_3^4(k,\kappa) + 256\gamma_0(k,\kappa) - 64\gamma_3(k,\kappa)\gamma_1(k,\kappa) + 16\gamma_3^2(k,\kappa)\gamma_2(k,\kappa)}{256}. \tag{3}$$

The eigenvectors corresponding to the energy eigenvalues $\in_j (s,\sigma,k,\kappa)$ are

$$|u^{(j)}(k,\kappa,z)\rangle = \varsigma_j^{-1/2}(k,\kappa)exp(-i\kappa z)\begin{pmatrix}\psi_1^j(k,\kappa)\\ \psi_2^j(k,\kappa)\\ \psi_3^j(k,\kappa)\\ \psi_4^j(k,\kappa)\end{pmatrix}, \; j=1,2,3,4,$$

$$\varsigma_j(k,\kappa) = |\psi_1^j|^2 + |\psi_2^j(k,\kappa)|^2 + |\psi_3^j(k,\kappa)|^2 + |\psi_4^j(k,\kappa)|^2$$

$$\psi_1^j = 1, \quad \psi_\nu^j(k,\kappa) = \Delta_\nu^{(j)}(k,\kappa)/\Delta^{(j)}(k,\kappa), \; \nu = 2,3,4. \qquad (4)$$

$$\Delta_2^{(j)}(k,\kappa) = -(\eta A_1 c\kappa)[((\eta A_1 c\kappa)^2 + (A_1 ak)^2) - \{(\in_j (s,\sigma,k,\kappa) - \epsilon(k,\kappa) + M)^2\}],$$

$$\Delta_3^{(j)}(k,\kappa) = (2M(\eta A_1 c\kappa)(A_1 ak_-)),$$

$$\Delta_4^{(j)}(k,\kappa) = ((\vartheta(k,\kappa) - M)^2 - (\in_j (s,\sigma,k,\kappa) - \epsilon(k,\kappa))^2 + (\eta A_1 c\kappa)^2 + (A_1 ak)^2) \times (A_1 ak_+),$$

$$\Delta^{(j)}(k,\kappa) = ((\in_j (s,\sigma,k,\kappa) - \epsilon(k,\kappa))^2 - M^2) \times \{\in_j (s,\sigma,k,\kappa) - \epsilon(k,\kappa) + M\} -$$

$$(\in_j (s,\sigma,k,\kappa) - \epsilon(k,\kappa) - M)\vartheta^2(k,\kappa) + (\in_j (s,\sigma,k,\kappa) - \epsilon(k,\kappa) - M)^2 \vartheta(k,\kappa) + ((\eta A_1 c\kappa)^2$$

$$+ (A_1 ak)^2) \times (\in_j (s,\sigma,k,\kappa) - \epsilon(k,\kappa) - M + \vartheta(k,\kappa)) + \vartheta^3(k,\kappa). \qquad (5)$$

These eigenvectors specify surface states. The wave number $\kappa$ is an unknown in Eq. (1) to (5). As already stated under OBC, we seek solution for this in the form $(\pm ib)$, $b > 0$, for ensuring an exponentially decaying term with plus sign for $z < 0$ (minus sign for $z > 0$) in the surface states. To determine an approximate value of $\kappa$ graphically we first write the energy eigenvalue equation given by the quartic above as an equation for $x = \kappa^2 = -b^2$ for a given energy eigenvalue $E_f$, at the $\Gamma$ (0,0) point. We obtain an octic in $\kappa$ given as $B_1 x^4 + Ax^3 + Bx^2 + C(M)x + D(M) = 0$. Here

$$A = [2(\eta A_1)^2 B_1^2 + (\eta A_1)^2 D_1^2 + B_1^2 D_1 E_f - 4D_1^3 E_f - 2\vartheta_0 B_1^3],$$

$$B = [(\eta A_1)^4 - 4(\eta A_1)^2 \vartheta_0 B_1 + 2(\eta A_1)^2 \epsilon_0 D_1 + 4(\eta A_1)^2 D_1 E_f + 3(2D_1^2 - B_1^2)E_f^2$$

$$- 12\epsilon_0 D_1^2 E_f + 4\epsilon_0 B_1^2 E_f - 8\vartheta_0 B_1 D_1 E_f + 6\vartheta_0^2 B_1^2],$$

$$C(M) = [4(\eta A_1)^2 \epsilon_0 E_f + (\epsilon_0^2 + M^2)(\eta A_1)^2 + 2(\eta A_1)^2 \vartheta_0^2 - 2\vartheta_0 B_1^3 - 2(\eta A_1)^2 E_f^2 + 2\epsilon_0 D_1 E_f^2$$

$$- 2\vartheta_0 B_1 E_f^2 - 4D_1 E_f^3 + 4D_1 E_f M^2 - 12\epsilon_0^2 D_1 E_f - 8\vartheta_0 B_1 \epsilon_0 E_f + 4\vartheta_0^2 D_1 E_f],$$

$$D(M) = [E_f^4 + \vartheta_0^4 - 2\vartheta_0^3 B_1 + 4\epsilon_0 E_f M^2 - 4\epsilon_0^3 E_f + 4\epsilon_0 \vartheta_0^2 E_f + 6\epsilon_0^2 E_f^2 - 12M^2 E_f^2$$

$$- 12\vartheta_0^2 E_f^2 - 4\epsilon_0 E_f^3]; \qquad (6)$$

the coefficients C(M) ( D(M) ) is increasing (decreasing)function of the exchange field M. Next, we plot $f_1 = B_1 x^4 + Ax^3 + Bx^2$ and $f_2 = C(M)x + D(M)$ as functions of the dimensionless number $x$ for a given M. In Figure 1 these plots are shown for $M = 0.10$, 0,30, 0.50, and 0.80. As the value of $M$ increases, the point of intersection of the two curves (which corresponds a solution sought for of the equation $B_1 x^4 + Ax^3 + Bx^2 + C(M)x + D(M) = 0$) shifts to the right. In Figure 2(e) we have shown a plot of $\exp(-b(\frac{z}{c}))$ as a function of (z/c) where $b \approx 1$ for $M = 0.5$. It is clear from the from Figure 1(e) that a thickness of the film (W/c) must be of O(10) to ensure surface state practically equals zero at $z = \pm \frac{W}{2}$. Since $c = 28.64$ Å, the thickness may be taken as $W \approx 30\ nm$ for $M = 0.5$ and $b \approx 1$ (see Figure 1(c)). Similarly, in order to have W/c ~O(10), for M = 0.3 and b = 1.1, we need to have the thickness $W \approx 27\ nm$. Thus, the exchange field from the magnetic dopants seems to be a factor in deciding the appropriate film thickness to ensure vanishing surface state. We now discuss the effect of the normal incidence of circularly polarized radiation field (CPRF) on the film. We shall assume the frequency of the incident light as $3 \times 10^{14}$ Hz and therefore the ratio $W/\lambda_{in} \approx 0.03 \ll 1$ (see Figure 2), where $\lambda_{in} \approx 1000\ nm$ is the wavelength of the incident radiation.

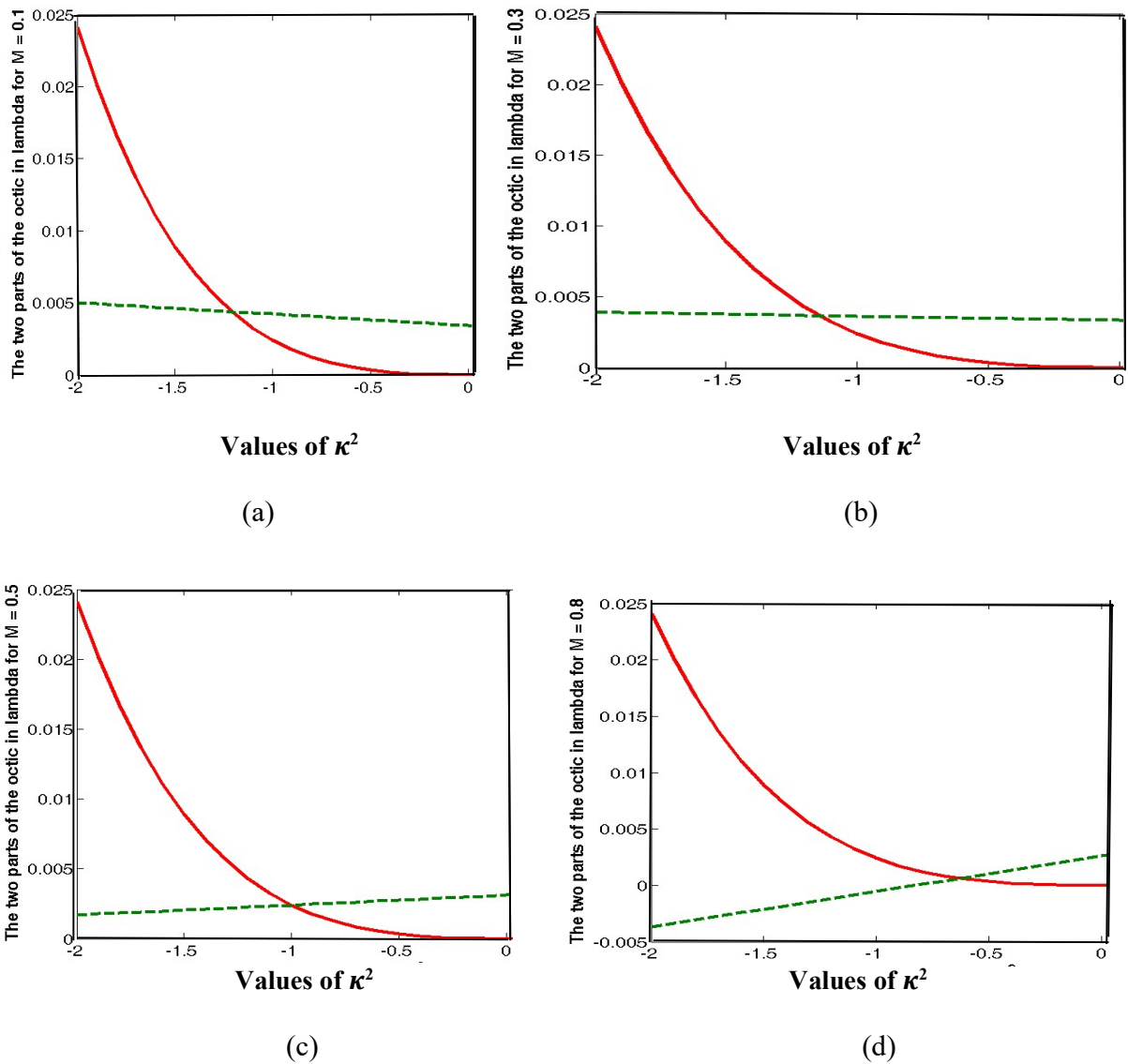

(a)

(b)

(c)

(d)

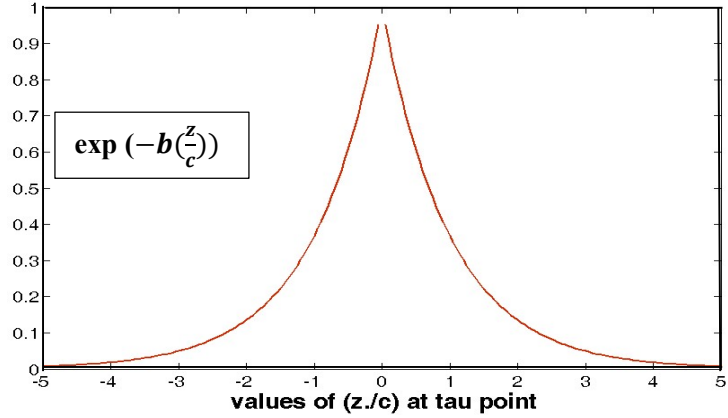

(e)

**Figure 1.** (a)-(d) The plots of $f_1 = B_1 x^4 + Ax^3 + Bx^2$ (in red ink) and $f_2 = C(M)x + D(M)$ (in green ink) as functions of the $(\lambda)^2$ for a given M. (a) M = 0.10 (b) M = 0,30 (c) M = 0.50 and (d) M = 0.80. The point of intersection of the two curves corresponds to a solution sought for of the equation $B_1 x^4 + Ax^3 + Bx^2 + C(M)x + D(M) = 0$). The numerical values of the parameters are $\epsilon_0 = -0.003, A_1 = 0.31, \eta = 0.16, B_1 = 0.18, B_2 = 1, E_f = 0.05, D_1 = 0.024$, and $D_2 = 0.34$. (e) A plot of $\exp(-b(\frac{z}{c}))$ as a function of $(z/c)$ where $b \approx 1$ for M = 0.5 (see Figure 2(c)).

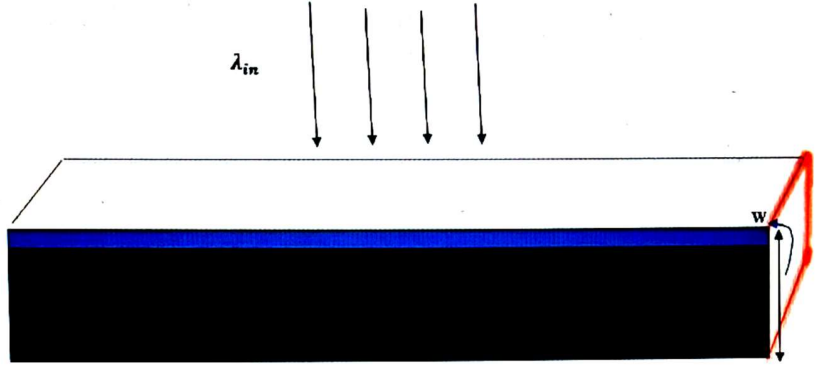

**Figure 2.** The circularly polarized radiation of wavelength $\lambda_{in}$ incident on the thin film of Bi$_2$Se$_3$ of thickness W where $\frac{W}{\lambda_{in}} \ll 1$.

Suppose the angular frequency of the optical Field incident on the film is ω. Taking this into consideration our Hamiltonian becomes time dependent. The Floquet theory can be applied to our time-periodic Hamiltonian $H_{surface}(t) = H_{surface}(t + T))$ with the period $T = 2\pi/\omega$. Then the wave function, in terms of the quasi-energies ε, has the form $\Psi(t) = \sum_\beta \exp\left(-i\left(\frac{\varepsilon}{\hbar} + \right.\right.$

$\beta\omega)t\big)\psi_\beta$ where $\beta$ is an integer. The element $H_{surface,\alpha,\beta}$ of the Hamiltonian is given by $\sum_\alpha H_{surface,\alpha,\beta}\psi_\alpha = \varepsilon\,\psi_\beta$, where $H_{surface,\alpha,\beta} = \alpha\hbar\omega\delta_{\alpha,\beta} + \frac{1}{T}\int_0^T H_{surface}(t)e^{i(\alpha-\beta)\omega t}dt$, where $(\alpha, \beta)$ are integers. This is the Floquet surface state Hamiltonian of the $Bi_2Se_3$ thin film. With $\alpha \neq \beta$, one can write the matrix as

$$H_{surface} = \begin{pmatrix} \cdots & \cdots & \cdots & \cdots & \cdots \\ \cdots & H_{surface,-1,-1} & H_{surface,-1,0} & H_{surface,-1,1} & \cdots \\ \cdots & H_{surface,0,-1} & H_{surface,0,0} & H_{surface,0,1} & \cdots \\ \cdots & H_{surface,1,-1} & H_{surface,1,0} & H_{surface,1,1} & \cdots \\ \cdots & \cdots & \cdots & \cdots & \cdots \end{pmatrix}. \quad (7)$$

The polarized light described by a time-varying gauge field $A(t) = A_0(\sin(\omega t), \sin(\omega t + \psi))$.

In particular, when the phase $\psi = \pi$ or 0, the optical field is linearly polarized. For $\psi = \pm\frac{\pi}{2}$, we have circularly polarized optical field (CPOF). When $\psi = -\frac{\pi}{2}$ ($\psi = +\frac{\pi}{2}$), the field is right-handed (left-handed) circularly polarized. Now once we have included a gauge field, it is necessary that we make the Peierls substitution $H_{surface}(t) = H_{surface}\left(\mathbf{k} - \frac{e}{\hbar}\mathbf{A}(t)\right)$. In view of the Floquet theory [12-17], our system now can be described by a time-independent effective Hamiltonian $H_{surface}^{effective}$ in the high-frequency limit:

$$H_{surface}^{effective} = \left(\widetilde{\epsilon(k,\kappa)}\sigma_0 + M\sigma_z\right)\otimes\tau_0 + \widetilde{\vartheta(k,\kappa)}\,\sigma_0\otimes\tau_z + \{\widetilde{A_1}(ak_x\sigma_x + ak_y\sigma_y)\eta\widetilde{A_1}c\kappa\sigma_z\}\otimes\tau_x, \quad (8)$$

where $\widetilde{\epsilon(k,\kappa)} = \epsilon(k,\kappa) + \alpha^2 A_0^2 D_2$, $\widetilde{\vartheta(k,\kappa)} = \vartheta(k,\kappa) - \alpha^2 A_0^2 B_2 \mp \left(\frac{a^2 A_0^2}{\hbar\omega}\right)A_1^2$, $\widetilde{A_1} = A_1\left(1\mp 2B_2\left(\frac{a^2 A_0^2}{\hbar\omega}\right)\right)$, $\epsilon(k,\kappa) = \epsilon_0 + D_1 c^2\kappa^2 + D_2 a^2 k^2$, $\alpha = \frac{ea\omega}{B_2}$, and $\vartheta(k,\kappa) = \vartheta_0 - B_1 c^2\kappa^2 - B_2 a^2 k^2$. The value of $\alpha^2 A_0^2$ (this quantity is the intensity of the radiation) is taken to be $0.65 - 0.90$ which is good for the near-infrared radiation field of frequency $\nu \sim 3\times 10^{14} Hz$ under consideration. Moreover, $+$ sign ($-$ sign) prefixed corresponds to the left-handed (right-handed) circularly polarized radiation. The eigenvalues ($E_j$) of this matrix is given by the quartic $E_j^4 + \gamma_{3F}(k,b)E_j^3 + \gamma_{2F}(k,b)E_j^2 + \gamma_{1F}(k,b)E_j + \gamma_{0F}(k,b) = 0$, where $\kappa^2$ has been replaced by $(-b^2)$, and

$$\gamma_{0F}(k,b) = (\eta A_1 c)^4 b^4 - 2(\eta A_1 c)^2((\widetilde{A_1}ak)^2 + \widetilde{\vartheta(k,b)}^2 - \widetilde{\epsilon(k,b)}^2 - M^2)\,b^2$$

$$+((\widetilde{A_1}ak)^2 + \widetilde{\vartheta(k,b)}^2)^2 - 2(\widetilde{A_1}ak)^2(\widetilde{\epsilon(k,b)}^2 - M^2),$$

$\gamma_{1F}(k,b) = 4(-(\eta A_1 c)^2 b^2 + (\widetilde{A_1}ak)^2)\widetilde{\epsilon(k,b)} + 4(\widetilde{\vartheta(k,b)}^2\widetilde{\epsilon(k,b)} - \widetilde{\epsilon(k,b)}^3 + \widetilde{\epsilon(k,b)}\,M^2),$
$\gamma_{2F}(k,b) = 2((\eta A_1 c)^2 b^2 - (\widetilde{A_1}ak)^2) - 2(\widetilde{\vartheta(k,b)}^2 - 3\widetilde{\epsilon(k,b)}^2 + M^2), \gamma_{3F}(k,b) - 4\widetilde{\epsilon(k,b)}.$
(9)

In view of the Ferrari's solution of a quartic equation, we find the roots of the quartic in $E_j$ similar to Eq. (2). In Figure 3 (a) we have plotted these energy eigenvalues $E_j$ as a function of ($ak$) for a given value of the intensity of incident radiation as 0.30. In Figure 3(b), however, the value of the intensity as 0.80. The value of the other parameters are $\epsilon_0 = -0.003$, $A_1 = 0.31$, $\eta = 0.16$, $B_1 = 0.18$, $B_2 = 1$, $M = 0.30$, $D_1 = 0.024$, $D_2 = 0.34$, and $\mu = 0$. Here $\mu$ is the chemical potential of the fermion number. We find that the band structure does not change much when M is increased from 0.3 up to 0.6.

The Kramer pair is degenerate when momentum +**k** becomes equivalent to −**k** due to the periodicity of the Brilloin zone (BZ), i.e., **k** + **G** = −**k** where **G** is a reciprocal lattice vector. Now, consider Figure 3(a). An inspection yields that here, particularly, the band $\in (s = -1, \sigma = +1, k, \lambda)$ possesses the above mentioned momentum **k** = (±2,0) /(0, ±2). The reason is that **k** (referred to as the time reversal invariant momentum − TRIM satisfies the condition **k** + **G** = −**k**. The vector **G** here is equal to (±4, 0) or (0, ±4). Let us now note that the Fermi energy $E_F \approx \mu = 0$ inside the gap intersects the closer surface state band $E_{31} = \in (s = -1, \sigma = +1, k, b)$ in the same BZ only once as the Kramer TRIM pair, whereas the band $E_{21} = \in (s = +1, \sigma = -1, k, b)$, which too is closer does not intersect. If there are odd pair of intersections, the surface state is topologically non-trivial (strong topological insulator), for disorder or correlations cannot remove pairs of such surface state crossings (SSC) by pushing the surface bands entirely above or below the Fermi energy $E_F$. When there are an even number of pairs of surface-state crossings, the surface states are topologically trivial (weak TI or conventional insulator), for disorder or correlations can remove pairs of such crossings. The inescapable conclusion is that the system under consideration is a strong topological insulator. Thus, the system qualifies to be called as quantum spin Hall (QSH) insulator. Interestingly, we find that the QSH state is possible even when the time reversal symmetry (TRS) is broken due to the finite value of the exchange field. The material band structures are usually characterized by Kane–Mele index $Z_2 = +1$ ($v_0 = 0$) and $Z_2 = -1$ ($v_0 = 1$). The former corresponds to weak TI, while the latter to strong TI. In Figure 3(c) and (d), we have shown the 3D plots of $E_{31} = \in (s = -1, \sigma = +1, k, b)$ as a function of the dimensionless wave number ($ak$) and $aA_0$ for $M = 0.3$. As the intensity is increased the system makes a crossover to QSH state (blue patch) starting from quantum anomalous Hall (QAH) region (red patch). Coming back to Figure 3(a), where the exchange field is M = 0.3 and $aA_0 = 0.60$, there is no TRIM pair. Thus, for this value of $aA_0$ the system is expected to be in quantum anomalous Hall (QAH) phase (see below).

The famous TKNN formula states that the Hall conductivity $\sigma_{xy}$ in a band insulator is related to the Chern number C by $\sigma_{xy} = \left(\frac{e^2 C}{2h}\right)$ and the Chern number C, which is an integer, is given by $C = 2 \int \int_{BZ} \Sigma_n \Omega_{xy}^{(n)}(k) \frac{d^2 k}{(2\pi)^2}$. Here BZ denotes an integral over the entire Brillouin zone (BZ), and the sum is over occupied bands. Suppose the eigenvectors corresponding to the energy eigenvalues of the Bloch Hamiltonian considered above are denoted by $u^{(n)}(k)$, where n is a band index. Suppose $u^{(n)}(k)$ is a Bloch state eigenfunction. The Berry curvature (BC) is defined as

$$\Omega_{xy}^{(n)}(k) = i \left\langle \partial_{k_x} u^{(n)}(k) \middle| \partial_{k_y} u^{(n)}(k) \right\rangle - i \left\langle \partial_{k_y} u^{(n)}(k) \middle| \partial_{k_x} u^{(n)}(k) \right\rangle \qquad (10)$$

Two important symmetries one usually considers here are time reversal symmetry (TRS) and inversion symmetry (IS). For the inversion symmetry (IS), $\Omega_{xy}^{(n)}(k) = \Omega_{xy}^{(n)}(-k)$ and, for the time reversal symmetry (TRS), $\Omega_{xy}^{(n)}(k) = -\Omega_{xy}^{(n)}(-k)$. Thus, for a system which is both TRS and IS compliant, $\Omega_{xy}^{(n)}(k) \equiv 0$. This means that in order to study the possible quantum anomalous Hall (QAH) effect, starting from the present Floquet Hamiltonian(8), we need to have broken TRS. This will ensure non-zero Berry curvature (BC). A non-zero BC is very much required to obtain anomalous Hall conductivity and to show that it is quantized in the case of an insulator. To obtain the Hall conductivity, calculation of the eigenvectors corresponding to the eigenvalues of the matrix $H^{surface}$ has been done. Next, the k-integration is necessary. For the integration purpose, we first divide the BZ into finite number of rectangular patches. We next determine the numerical values corresponding to each of these patches of the momentum-dependent density and sum these values. The sum is then divided by the number of patches. We have generated these values through a surface plot. We find, for example, C = 0.7523 for M = 0.3. For the calculation of the Hall conductivity in this phase, we have considered two bands, viz. $\in (s = -1, \sigma = +1, k, \lambda)$ and $\in (s = +1, \sigma = -1, k, \lambda)$ close to the Fermi energy $E_F$. There is a caveat though, which concerns the validity of the calculated result. The Hall conductivity cannot be determined as such from the 2D Dirac model since the former requires an integral over the whole BZ, as we have seen above. The integral is outside the Dirac model's range of validity. To circumvent the problem we can choose a momentum space cut-off small compared to the size of BZ and large enough to capture nearly all the Berry curvature integral. This will be within the range of validity of the 2D Dirac model. In this limit, the Hall conductivity $\sigma_{xy} \rightarrow \left(\frac{e^2 C}{2h}\right)$ where C $\rightarrow$ an integer.

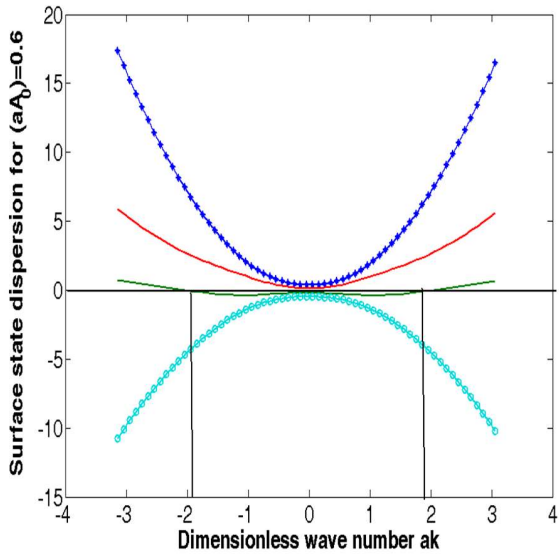
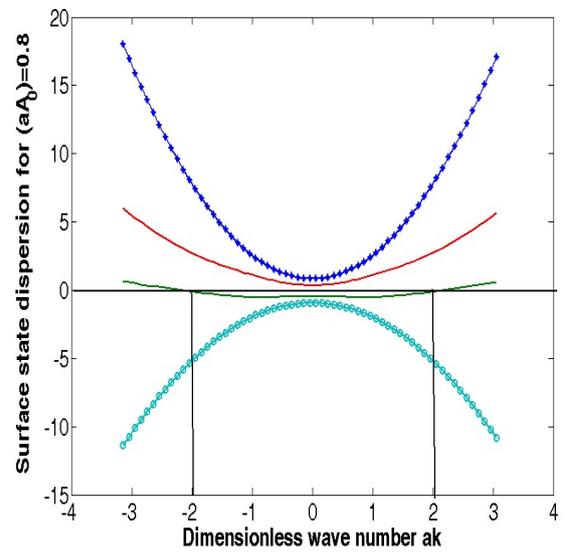

(a)                                           (b)

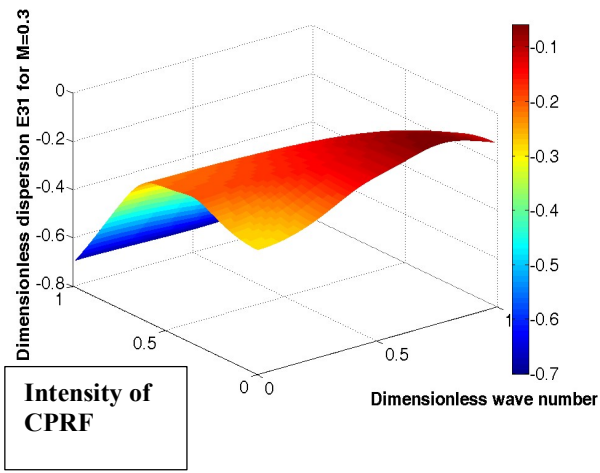
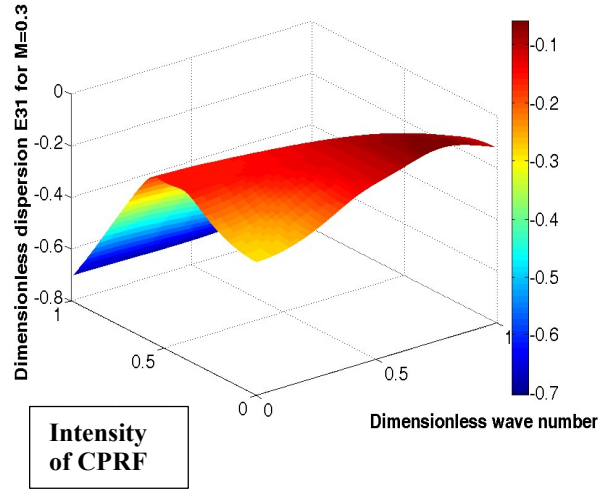

(c) Left handed CPRF                       (d) Right handed CPRF

**Figure 3. (a), and (b)** The plots of $E_j$ as functions of the dimensionless wave number $(ak)$ for a given intensity of incident radiation. The numerical values of the other parameters are $\epsilon_0 = -0.003, A_1 = 0.31, \eta = 0.16, B_1 = 0.18, B_2 = 1, E_f = 0.05, D_1 = 0.024, \mu = 0, M = 0.30, b = 0.98,$ and $D_2 = 0.34$. The Fermi energy $E_F = 0$ is represented by a horizontal line. **(c), and (d)** The 3D plots of $E_{31} = \in (s = -1, \sigma = +1, k, \lambda)$ as a function of the dimensionless wave number $(ak)$ and the intensity of the incident radiation (left handed as well as right handed CPRF). As the latter is increased the system makes a crossover to QSH state (blue patch) starting from quantum anomalous Hall (QAH) state (red patch).

The light–matter interaction of topological insulator material like Bismuth Selenide is found to be a desirable system for the investigation of linear and nonlinear optical properties. This has been attracting considerable amount of research interest lately **[18 - 21]**. The present work is motivated by these investigations. We have derived here an effective Hamiltonian for the surface states of a topological insulator thin film incorporating the effect of the normal incidence of CPRF on the film using the Floquet theory in the high-frequency limit. We found that the surface of the system has states, which come in an odd number of Kramers' doublets when intensity of radiation attains a critical value. In fact, the graphical representations lead us to the conclusion that the system considered here makes a crossover to QSH state from QAH state when the intensity of incident radiation $\gtrsim 0.6$ with the exchange field strength M $\neq 0$. In the former phase, the surface comprises of 'helical liquids', which (helicity) is one of the most unique properties of a topologically protected surface. The question "whether this crossover is a phase transition" could only be settled through thermodynamic consideration.

Finally, we note that the surface band gap $(E_{31} - E_{21})$ could be controlled by changing the parameter $b$. We have shown above that the exchange field strength is a factor to decide the value of the parameter $b$, and consequently the appropriate film thickness W, to ensure vanishing surface state at $z = \pm\frac{W}{2}$. Tunable band gap provides a promising platform to realize optical devices for room temperature application **[22,23]**. We anticipate that our theoretical result will be useful for future experiments.